\documentclass[10pt, conference]{IEEEtran}

\usepackage{multirow}
\usepackage{todonotes}
\usepackage{hyperref}
\usepackage{cite}
\usepackage{amsmath,amssymb,amsfonts}
\usepackage{graphicx}
\usepackage{dcolumn}
\usepackage{bm}
\usepackage{ulem}
\usepackage{listings}
\usepackage{xcolor}
\usepackage{cuted}
\usepackage{braket}
\usepackage{caption}

\usepackage{duckuments}
\usepackage{tikz}

\title{Symbolic Hamiltonian Compiler for Hybrid Qubit-Boson Processors}

\author{
    Ethan Decker$^{1}$,
    Erik Gustafson$^{2}$,
    Evan McKinney$^{3}$,
    Alex K. Jones$^{4}$,\\
    Lucas Goetz$^{5}$,
    Ang Li$^{6}$,
    Alexander Schuckert$^{7}$,
    Samuel Stein$^{6}$,
    Gushu Li$^{1}$,
    Eleanor Crane$^{8,9}$\\
    $^{1}$University of Pennsylvania, Philadelphia, PA, USA\\
    $^{2}$Universities Space Research Association, RIACS at NASA Ames Research Center, Mountain View, CA, USA\\
    $^{3}$University of Pittsburgh, Pittsburgh, PA, USA\\
    $^{4}$Syracuse University, Syracuse, NY, USA\\
    $^{5}$ETH Zurich, Zurich, Switzerland\\
    $^{6}$Pacific Northwest National Laboratory, Richland, WA, USA\\
    $^{7}$University of Maryland, College Park, MD, USA\\
    $^{8}$Massachusetts Institute of Technology, Cambridge, MA, USA\\
    $^{9}$King's College London, London, UK
}

\begin{document}

\maketitle

\begin{abstract} 
Quantum simulation of the interactions of fermions and bosons -- the fundamental particles of nature -- is essential for modeling complex quantum systems in material science, chemistry and high-energy physics and has been proposed as a promising application of fermion-boson quantum computers, which overcome the overhead encountered in mapping fermions and bosons to qubits. However, compiling the simulation of specific fermion-boson Hamiltonians into the natively available fermion-boson gate set is challenging. In particular, the large local dimension of bosons renders matrix-based compilation methods, as used for qubits and in existing tools such as Bosonic Qiskit or OpenFermion, challenging.  We overcome this issue by introducing a novel symbolic compiler based on matrix-free symbolic manipulation of second quantised Hamiltonians, which automates the decomposition of fermion-boson second quantized problems into qubit-boson instruction set architectures. This integration establishes a comprehensive pipeline for simulating quantum systems on emerging qubit-boson and fermion-boson hardware, paving the way for their large-scale usage.
\end{abstract}


\section{Introduction}
Quantum computers are emerging as a promising paradigm to overcome the limitations of traditional classical computing in specific areas, such as material science~\cite{Cao_2019,TILLY20221,Sawaya:2023nmv}, simulating high energy and nuclear physics~\cite{Bauer:2022hpo,Alam:2022crs} as well as non-quantum applications such as classical optimization problems~\cite{blekos2024review}. Many of the promising applications relate to the simulation of the fundamental particles of nature --fermions, for example electrons, and bosons, for example the vibrations of a molecule. In order to simulate such problems on a qubit quantum computer, fermions and bosons need to be mapped to qubits. While a single fermion site can be mapped to a single qubit, bosons are $n$-level systems which require at least $\log_2(n)$ qubits to realize, and in addition a large gate overhead when simulating time evolution~\cite{crane2024}.

To avoid this overhead, recently qubit-boson architectures have been developed which directly manipulate bosonic modes in addition to qubits~\cite{crane2024, liu2024hybrid}. In ~\cite{crane2024} examples are provided of the quantum simulation of fermions, bosons and gauge fields in $1$D and $2$D with qubit-boson architectures, as will be discussed in this work. Their advantage over all-qubit architectures was demonstrated by performing an end-to-end comparison of the gate complexity for the gauge-invariant hopping term of a $Z_2$ lattice gauge theory, for which there was an improvement of the asymptotic scaling with the boson number cutoff $S$ from $O(\log(S)^2)$ to $O(1)$ and for bosonic matter a constant factor improvement of better than $10^4$, as well as an improvement from $O(\log(S))$ to $O(1)$ for the U(1) lattice gauge theory magnetic field term.

Similarly, recently fermion-qubit platforms have been proposed~\cite{gonzalez-cuadra_fermionic_2023} and their fault-tolerance constructed and advantages calculated in Ref.~\cite{Schuckert_2024}. 

These new platforms with computational fermions and bosons are ideally positioned to accelerate the simulation of quantum processes in nature involving fermions and bosons, such as material science, quantum chemistry and nuclear physics.

In order to use these boson-fermion quantum computers, a target problem needs to be converted into a series of operations on the quantum computer, i.e. gates. The general problem of converting an arbitrary n-qubit unitary into a specific gate set is exponentially hard\cite{dawson2005solovaykitaevalgorithm}, therefore more efficient problem-taylored approaches such as Trotterization~\cite{childs_theory_2021} are warranted.

Existing software development kits (SDKs) tend to focus on either high-level abstraction or low-level gate compilation, often requiring substantial domain expertise to navigate, which limits the broader adoption of quantum technologies~\cite{mcclean2019openfermionelectronicstructurepackage, qiskit2024, Killoran_2019}. Compounding this challenge is the rapidly evolving quantum hardware landscape, where diverse platforms such as qubit-boson hardware 
each demand tailored, resource-intensive compilation strategies~\cite{bernardini2023quantumcomputingtrappedions, google2023suppressing, bluvstein2024logical}.
\begin{figure}[!th]
\includegraphics[width=\linewidth]{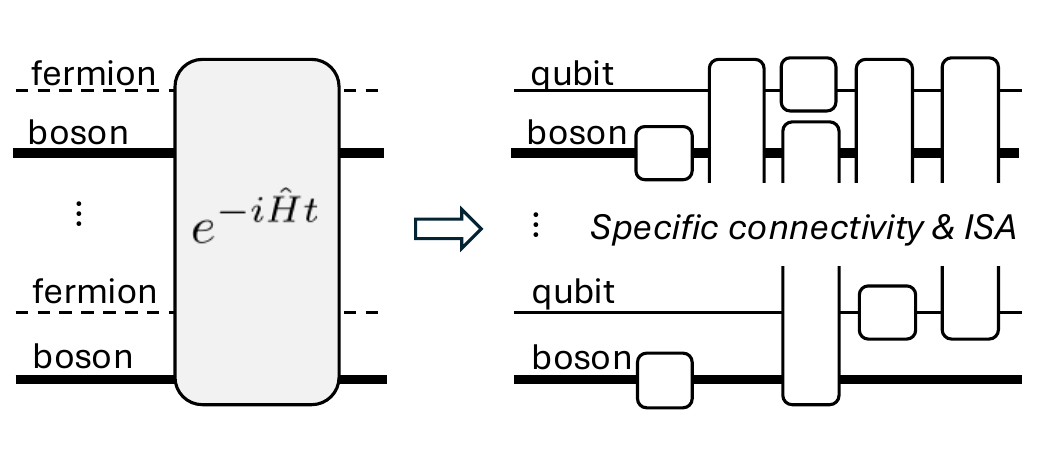}
\caption{Our software package finds the decomposition of the time evolution under a target fermion-boson Hamiltonian into an instruction set architecture (ISA) and connectivity provided by qubit-boson hardware. This way, it provides a method to decouple the domain expert scientist who is interested in simulating Hamiltonians from the hardware software expertise familiar with ISAs and connectivities.}
\label{fig:ual}
\end{figure}

Flagship SDKs exemplify current progress in quantum software development~\cite{qiskit2024,Killoran_2019,cirq_developers_2024_11398048}. Although tools like Qiskit streamline gate-level programming, users often need to manually translate high-level unitaries into device-specific sets of one- and two-qubit gates~\cite{crane2024},~\cite{PhysRevA.97.062311},~\cite{liu2024hybrid}, despite the availability of some community-developed aids that work to simplify this process~\cite{javadiabhari2024quantumcomputingqiskit},~\cite{mcclean2019openfermionelectronicstructurepackage}, \cite{decker2025kernpilercompileroptimizationquantum}. 
This manual decomposition exposes hardware details at the software level, relying on expert knowledge of the underlying physics and forcing programmers to approximate complex Hamiltonians—a task compounded by the constraints of error correction protocols that often enforce discrete gate sets. Together, these demands create a steep learning curve and hinder broader accessibility, as efficiently programming quantum devices increasingly depends on intricate domain-specific expertise rather than purely algorithmic skill.

In ~\cite{crane2024}, fermion-boson Hamiltonian simulation problems were compiled to qubit-boson circuits by hand, developing along the way a number of compilation rules. In order to code them into a qubit-boson device, a wrapper for Qiskit called \textbf{Bosonic Qiskit}~\cite{Stavenger:2022wzz} was constructed. However, \textbf{Bosonic Qiskit}~\cite{Stavenger:2022wzz} enables the assembly of qubit-boson gates but does not directly apply compilation rules to second quantized Hamiltonians, and neither do \textbf{MQT Qudits}~\cite{Mato:2024hnd} which is a useful software for qudit circuits. Therefore, in these solutions, hybrid Hamiltonian terms and fermionic statistics must be encoded manually as the abstraction remains at the gate level. Other approaches, such as the \textbf{SimuQ} framework~\cite{Peng:2023fmd}, elevate the level of abstraction by allowing the expression of Hamiltonians using novel languages. While this represents progress, SimuQ's applicability is limited to spin models and analog quantum computation (a less general version than gate based quantum computation). Consequently, it lacks the generality required for gate-based simulations of algorithms like Shor's or Grover's, which are essential for demonstrating quantum advantage. \textbf{Bosehedral}~\cite{zhou:2024bosehedral} focuses on optimizing the decomposition of linear interferometers and thus does not support the compilation of generic spin-boson Hamiltonians.

In response to these limitations, we introduce a vertically integrated SDK that fully decomposes time evolution under a fermion-boson Hamiltonian into weight one and two boson-fermion-qubit unitaries, integrating prior theoretical compilation work~\cite{crane2024, Schuckert_2024, kang2025, liu2024hybrid}. Our solution accepts general combinations of Hamiltonians or unitaries and an input ISA, producing quantum machine code that can run on any quantum computer adhering to that ISA. The core of our method processes Hamiltonians symbolically to avoid exponential data structures. We employ a pipeline of equivalences--many of which are exact--to systematically factor high-weight unitaries into sequences of weight-one and weight-two unitaries. Building on OpenFermion~\cite{mcclean2019openfermionelectronicstructurepackage}, our interface expresses arbitrary unitaries in terms of spin, quadrature, and second quantization operators. By introducing a unique normal-ordering scheme, we ensure unambiguous processing and straightforward identification of symbolic equivalences.

In summary, our main contributions are:
\begin{enumerate}
    \item An automated set of compiler stages to go from high level unitaries defined by a boson-fermion Hamiltonian into gate level quantum instructions for qubit-boson hardware.
    \item In most cases, our compiler achieves a number of gates which is independent of the boson cutoff, as opposed to the quadratic scaling in state-of-the-art qubit-only approaches
    \item The runtime of the compiler scales polynomially with the number of sites, with small exponents.
\end{enumerate}

\section{Technical Background} 

\label{sec:technicalstuff}
\subsection{The second quantization of quantum mechanics for hamiltonians}
In the framework of second quantization, ladder operators serve as fundamental tools for describing and manipulating quantum systems composed of indistinguishable particles. These operators, commonly denoted as $a$ (annihilation operator) and $a^\dagger$ (creation operator), are tailored to the statistics of the particles under consideration---bosons or fermions.

For bosonic systems, such as photons or atoms, the ladder operators adhere to the commutation relation:

\begin{equation}
[\hat{a}_i,\hat{a}^\dagger_j] \equiv \hat{a}_i\hat{a}^\dagger_j - \hat{a}^\dagger_j \hat{a}_i = \delta_{i,j},
\end{equation}
where $i$ and $j$ denote different sites of the system. This relation reflects the symmetric nature of bosonic wavefunctions. In addition, bosons can occupy the same quantum state simultaneously, i.e. by contrast to qubits for which each site can be either in state $\ket{0}$ or $\ket{1}$, bosons can also be in state $\ket{2}$, $\ket{3}$, up to in principle $\ket{\infty}$.

Conversely, for fermionic systems, which include electrons and protons that obey the Pauli exclusion principle, the operators (denoted as $\hat c$ and $\hat c^\dagger$) satisfy the anticommutation relation:

\begin{equation}
\{\hat{c}_i,\hat{c}^\dagger_j\} \equiv \hat{c}_i\hat{c}^\dagger_j + \hat{c}^\dagger_j \hat{c}_i = \delta_{i,j}.
\end{equation}
This anticommutation relation ensures that no two fermions can occupy the same quantum state, capturing the antisymmetric nature of fermionic wavefunctions. Therefore, a fermionic site can either be in state $\ket{0}$, indicating the absence of a fermion and $\ket{1}$, indicating its presence. This distinction is crucial for accurately modeling different quantum systems, as it directly affects the allowed configurations and statistical behavior of the particles involved.

Moreover, in the context of bosonic systems, ladder operators are intrinsically related to the quadrature operators---the position $x$ and momentum $p$---which are fundamental observables in quantum mechanics. Specifically, the ladder operators can be expressed in terms of these quadrature operators as:

\begin{equation}
\hat{a}_j = \frac{1}{\sqrt{2}}(\hat{x}_j + i\hat{p}_j), \quad \hat{a}^\dagger_j = \frac{1}{\sqrt{2}}(\hat{x}_j - i\hat{p}_j),
\end{equation}

where we have set $\hbar = 1$ for simplicity. This relationship highlights how ladder operators encapsulate both the position and momentum information of the quantum state. The quadrature operators themselves are Hermitian, corresponding to measurable physical quantities, while the ladder operators are non-Hermitian and facilitate transitions between quantum states by lowering or raising the energy level by one quantum.

\subsection{Trotterization and Baker-Campbell-Hausdorff formulas}

At the core of our simulation capabilities are two fundamental quantum simulation kernels: the \textbf{Trotterization algorithm} and the \textbf{Baker-Campbell-Hausdorff (BCH) formula}~\cite{Sefi_2011}. 
The Trotterization algorithm~\cite{childs_theory_2021} is ubiquitous in quantum compilation because it allows us to decompose the exponential of a sum of non-commuting operators into a product of exponentials of individual terms. 
This decomposition is formalized by the Trotter-Suzuki approximation

\begin{equation}
e^{-i\hat{H}t} = \left( \prod_{k} e^{-i\hat{H}_k t / n} \right)^n + \mathcal{O}(t^2),
\end{equation}
where $H= \sum_{k} H_k$ and $n$ is the number of Trotter steps. Higher-order approximations of the Trotter-Suzuki formula exist, offering faster convergence and improved accuracy by reducing the error terms associated with the approximation.

The \textbf{BCH formula} provides a method to combine exponentials of non-commuting operators, which is particularly useful for handling high-weight terms in the Hamiltonian that are challenging to decompose~\cite{crane2024, kang2025, crane2021}. The adapted version of the BCH formula we employ is given by:

\begin{equation}
e^{t\hat{A}} e^{t\hat{B}} = e^{t(\hat{A} + \hat{B}) + \frac{t^2}{2}[\hat{A}, \hat{B}]} + \mathcal{O}(t^3).
\end{equation}

To translate fermionic systems into qubit-based representations compatible with quantum hardware, we implement \textbf{fermion-to-qubit mappings}, specifically leveraging the \textbf{Jordan-Wigner (JW) transformation}, with Hemery et al.~\cite{fermi-hubbard} performing the largest digital simulation of the spinful Fermi-Hubbard model to date (see also \cite{Nigmatullin:2024mbq} for a large simulation of the spinless model), which compiles the fermions using this mapping in the form of Fermi-Swap networks~\cite{kivlichan_quantum_2018}. This mapping converts fermionic creation and annihilation operators into qubit operators while preserving the essential anticommutation relations:

\begin{figure}
    \centering
    \includegraphics[width=0.8\linewidth]{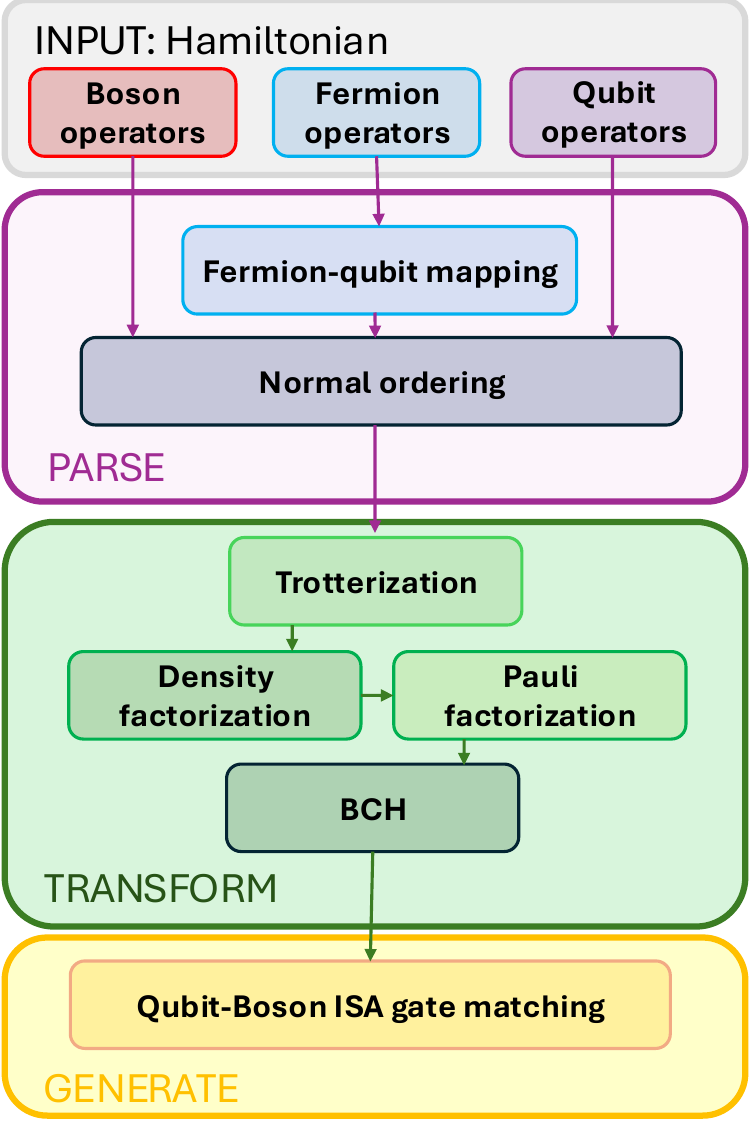}
    \caption{Pipeline of our qubit-boson compiler. The parse section breaks down the Hamiltonian terms into various symbolic components. The transformations break the Hamiltonians into factorized terms following \cite{crane2024}. The generate section maps the finalized transformation to a target ISA.}
    \label{fig:flowchart}
\end{figure}

\begin{equation}
\hat c_j^\dagger = \left( \prod_{k=1}^{j-1} \hat Z_k \right) \hat \sigma_j^+, \quad
\hat c_j = \left( \prod_{k=1}^{j-1} \hat Z_k \right) \hat \sigma_j^-,
\end{equation}

where $\sigma_j^\pm$ are the raising and lowering operators for the $j$-th qubit, and $Z_k$ is the Pauli-$Z$ operator acting on the $k$-th qubit. Preserving these commutation relations is crucial for accurately simulating fermionic behavior on quantum computers. Our compiler's modular design allows for the incorporation of alternative mappings, such as the Bravyi-Kitaev \cite{bravyi2002fermionic} or the parity mapping \cite{seeley2012bravyi}, which may offer optimized performance for specific types of simulations. By accommodating various mappings, we ensure that our compiler remains adaptable and efficient across a broad spectrum of quantum simulations, further decoupling the programmer from the intricacies of quantum hardware.

\subsection{Qubit-Boson Instruction Set Architecture}
We show the most frequently used qubit-boson gates in Table ~\ref{tab:gates}.
    \begin{table}
    \def\arraystretch{1.5}
    \centering
    \captionsetup{labelfont=normalfont, textfont=normalfont}
    \caption{Frequently used qubit-boson gates, from ~\cite{Stavenger:2022wzz}. \label{tab:gates}}
    \begin{tabular}{c|c}
        \hline\hline
        Name & Operation\\\hline \hline
        \multicolumn{2}{c}{Bosonic Operators}\\\hline \hline
        $\mathrm{R}_i(\theta)$ & $e^{-i\theta \hat{n}_i}$\\\hline
        $\mathrm{D}_i(\alpha)$ & $e^{\alpha \hat{a}^\dagger_i - \alpha^*\hat{a}_i}$\\\hline
        $\mathrm{BS}_{i,j}(\phi,\theta)$ & $e^{-i\theta(e^{i\phi}\hat{a}^{\dagger}_i\hat{a}_j + e^{-i\phi}\hat{a}_i\hat{a}_j^{\dagger})}$\\\hline \hline
        \multicolumn{2}{c}{Qubit Operators}\\\hline \hline
        $R_j^z(\theta)$ & $e^{-i\theta \hat{Z}_j}$\\\hline
        $R_j^y(\theta)$ & $e^{-i\theta \hat{Y}_j}$\\\hline
        $R_j^x(\theta)$ & $e^{-i\theta \hat{X}_j}$\\\hline
        CNOT & Controlled X \\\hline \hline
        \multicolumn{2}{c}{Coupled Bosonic-Qubit Operators}\\\hline \hline
        CR$_{i,j}(\theta)$ & $e^{-i \theta/2 \hat{Z}_i \hat{n}_j}$
        \\ \hline
        C$\Pi_{i,j}$ & $e^{-i\pi/2 \hat{Z}_i \hat{n}_j}$
        \\ \hline
        SNAP$_{i,j}(\vec{\theta})$ & $e^{-i\hat{Z}_i\sum_n \theta_n |n\rangle\langle n|}$\\\hline
        SQR$_{i,j}(\vec{\theta}, \vec{\phi})$ & $\sum_{n}\hat{R}_{i}^{\phi_n}(\theta_n)|n\rangle\langle n|_j$\\\hline\hline
    \end{tabular}
\end{table}
    
    \section{Compiler Technical Discussion} 

Our compiler's \textbf{intermediate representation (IR)}, a rewritten version of the Hamiltonian for computational tractability, and the \textbf{instruction set architecture (ISA)} use OpenFermion's backend, representing fundamental operators as tuples $(\text{action}, \text{index})$. Here, $\text{action}$ specifies the operator type (e.g., creation, annihilation, or Pauli $\hat{X}$, $\hat{Y}$, $\hat{Z}$), and $\text{index}$ labels the site it acts upon. Complex Hamiltonian terms are tuples of such fundamental operators paired with scalar coefficients.

The compiler is based on the compilation rules defined in Tables 2, 4 and 5 of Ref.~\cite{crane2024}. In the following, we walk through Fig.~\ref{fig:flowchart} and define each stage.
    
\subsection{Input: Hamiltonian}
    Our framework does not rely on directly constructing the Hamiltonian matrix and is therefore not restricted in size. This is crucial as for bosonic systems, such a construction is in principle impossible due to the infinite local Hilbert space dimension of a single bosonic site.  The user simply defines the Hamiltonian to be time-evolved under in second quantisation and our package then compiles time evolution under the Hamiltonian into weight-1 and weight-2 gates (i.e. the qubit-boson generalisations of single-qubit and two-qubit gates).
    
    We provide an example in Fig.\,\ref{fig:examplehamcode} for implementing the following Hamiltonian
\begin{align}
    &\hat H_{H.H.} = \sum_{i}\sum_{\sigma\in \lbrace \uparrow,\downarrow\rbrace } (\hat{c}_{i,\sigma}^{\dagger}\hat c_{i+1,\sigma} + \hat{c}^{\dagger}_{i+1,\sigma}\hat{c}_{i,\sigma}) + \sum_{i}^{N_\mathrm{s}}\omega_i \hat{b}_i^{\dagger}\hat{b}_i\notag\\
    &+\sum_{i}^{N_\mathrm{s}} U_i \hat{c}^{\dagger}_{i,\uparrow}\hat{c}_{i,\uparrow}\hat{c}^{\dagger}_{i,\downarrow}\hat{c}_{i,\downarrow}+g\sum_{i}^{N_\mathrm{s}}\sum_{\sigma\in\lbrace \uparrow,\downarrow\rbrace}\hat{c}_{i,\sigma}^{\dagger}\hat{c}_{i,\sigma}(\hat{b}_{i}^{\dagger} + \hat{b}_{i}).\label{eq_HH},
\end{align}
which is the Hubbard-Holstein Hamiltonian appropriate to simulating phonons coupled to electrons.
    
    \begin{figure}[!ht]
    \begin{lstlisting}
# Import symbolic compiler extension 
from qubit_boson_compiler import SCompile

# Import bosonic operators
from openfermion_hybrid import c,a

# Assign values to the Hamiltonian parameters
num_sites = 3 # number of sites in the model
t = 1 # fermionic hopping parameter
g = 1 # hybrid hopping parameter
U = 0.6 # fermionic onsite parameter
omega = 1 #bosonic density parameter 
dt = 0.1 # define a time slice

# Define Hamiltonian
def create_hubbard_holstein_hamiltonian_NN(num_sites, t, g, U):
    # Create empty operator object
    H = c() 
    
    for i in range(num_sites):
        j = (i + 1) # nearest neighbor labelling
        # Create fermionic hopping term
        if j <= num_sites:    
                H += t*c(f"{i}{spin}^ {j}{spin}") 
                # Add Hermitian conjugate
                H += t*c(f"{j}{spin}^ {i}{spin}")       
        # Create bosonic density term
        H += omega*a(f"{i}^ {i}")
        # Create onsite density-density term 
        for spin in [u,d]: # up, down
            H+= U*c(f"{j}{spin}^ {i}{spin}")*c(f"{j}{spin}^ {j}{spin}")
        # Create hybrid interaction term
        for spin in [u,d]: # up, down 
            H=g*c(f"{j}{spin}^ {i}{spin}")*a((f"{i}^")+a(f"{i}"))
    # Return the Hamiltonian with the bosonic Hilbert space truncated at Fock state 5
    return convertAlgorithm(dt * H, fockspace=5)

# run the function to create the Hamiltonian
circuit=create_hubbard_holstein_hamiltonian_NN(num_sites,t,omega,U) 
# Compile using the qubit_boson_compiler
SCompile(circuit) 

    
    \end{lstlisting}
    \caption{\textbf{Example code.} We define a Hubbard-Holstein Hamiltonian provided in Eq. (\ref{eq_HH}) and then compile subsequently into the gateset of Table \ref{tab:gates}. Here c() is a fermionic operator and a() is a bosonic operator.}
    \label{fig:examplehamcode}
    \end{figure}

    \subsection{Parse}
    Our qubit-boson compiler comprises two critical parsing stages—\textbf{normal ordering} and \textbf{fermion-to-qubit mapping}—that serve as intermediaries between the user interface and the compiler backend. These stages enable a simplification of the subsequent compilation pipeline.
    \subsubsection{Fermion-to-qubit Mapping}
    The fermionic operators must be mapped to qubits in order to be executed on a qubit-boson device because of the lack of native fermionic degrees of freedom (as were proposed for example here~\cite{gonzalez-cuadra_fermionic_2023,Schuckert_2024}). The \textbf{fermion-to-qubit mapping} stage is automatically executed when users declare fermionic operators. For simplicity and computational efficiency, we default to the general-purpose Jordan-Wigner (JW) mapping, though more choices of mappings exist~\cite{bravyi2002fermionic} for more specific applications.
    
    \subsubsection{Normal Ordering}
    In the \textbf{normal ordering} stage, which we generalized from OpenFermion~\cite{mcclean2019openfermionelectronicstructurepackage}'s to account for mixed bosonic-fermionic matter, operators are arranged such that number operators appear leftmost, followed by qubit operators, and finally ladder operators. This ordering matches our decomposition strategy, inspired from~\cite{crane2024}, since we first handle number operators, then apply exact equivalences for qubit operators, and only then address complex combinations of ladder operators using the Baker-Campbell-Hausdorff (BCH) formula.

    While our normal ordering algorithm can be exponential in the length of the operator weight in the worst case, realistic many-body Hamiltonians only consist of terms which have weight $k$, where $k=\mathcal{O}(1)$ in $N_\mathrm{s}$. Moreover, since bosons and qubits commute within their respective groups, identifying number operators and performing fermion-to-qubit mapping before normal ordering generally streamlines the process, making this step efficient in practice.
    
    When dealing with fermionic operators on overlapping modes, appropriate minus signs arise whenever operators are swapped. In bosonic or non-overlapping fermionic modes, no sign changes are necessary.

    As an example for normal ordering, consider the following operator acting on two modes (labeled 1 and 2):
\[
\hat{O} \;=\; \hat{a}_1 \, \hat{X}_2 \,\bigl(\hat{a}_2^\dagger \hat{a}_2\bigr) \, a_1^\dagger.
\]
Normal ordering then amounts to rewriting this operator as

\[
\begin{aligned}
\hat{O} = \underbrace{(\hat{a}_2^\dagger \hat{a}_2)}_{\text{number}}
   \;\; \underbrace{\hat{X}_2}_{\text{qubit}}
   \;\; \underbrace{\hat{a}_1\, \hat{a}_1^\dagger}_{\text{ladder}}
   \;\; .
\end{aligned}
\]

\subsection{Transform}

In the transformation stage, we break the Hamiltonian into factorized terms. It is organized in a directed acyclic graph (DAG), enabling systematic transformations via node manipulations and reduces arbitrary polynomial expressions in bosonic and spin degrees of freedom into the overcomplete ISA defined in Tab.~\ref{tab:gates} which contains only weight-one and weight-two unitaries.
    
The compilation pipeline consists of three major passes—\textbf{Trotterization}, \textbf{Factorization}, and \textbf{Baker-Campbell-Housdorf (BCH) Expansion}—each building upon the structured IR generated by the preprocessing stages.

These methods are summarized in Tab.~\ref{tab_rules} which is reproduced with permission from \cite{crane2021}.

\begin{table*}[]
    \centering
    \tabcolsep=0.2cm
    \setlength\extrarowheight{5pt}
    \begin{tabular}{|c|c|c|c|}
    \hline
         Gate & Expression & Ancilla qubit & Gate Decomposition
         \\[5pt] \hline  \hline
         Density factorization
         & $\exp{(-i(\hat n_i\hat O_j))}$ & Required
     & $\prod_{k=0}^{K-1} \operatorname{SQR}_{\textrm{anc},i}(\vec{\pi}_{k}, \vec{0}) e^{-i2^{k-1}\hat O_j}e^{i2^{k-1}\hat{Z}_{\textrm{anc}}\hat O_j}\operatorname{SQR}_{\textrm{anc},i}(-\vec{\pi}_{k}, \vec{0})$
         \\[5pt] \hline \hline
         Pauli factorization & $\exp{\left(\hat Z_i \left(\hat{\Theta} \hat{a}_j^\dagger - \hat{\Theta}^{\dagger} \hat{a}_j\right)\right)}$ & None & $\operatorname{C\Pi}_{i,j}e^{i\left(\hat{\Theta} \hat{a}_j^\dagger + \hat{\Theta}^{\dagger} \hat{a}_j\right)}\operatorname{C\Pi}_{i,j}^{\dagger}$
         \\[5pt] \hline \hline
          BCH & $\exp\left(-2i (\hat{O}_{\vec{I}}\hat{O}_{\vec{J}})\theta^2\right)$  & Required & $\operatorname{CU}^X_{k,\vec{I}}(\theta)\operatorname{CU}^Y_{k,\vec{J}}(\theta)\operatorname{CU}^X_{k,\vec{I}}(-\theta)\operatorname{CU}^Y_{k,\vec{J}}(-\theta)+\mathcal{O}(\theta^3)$
          \\[5pt] \hline \hline
          Trotter & $\exp{\left(-i \left(\hat{O}_{\vec{I}}+\hat{O}_{\vec{J}}\right)\theta\right)}$ & Required & $\operatorname{CU}_{i,\vec{I}}^{Z}(\theta)\operatorname{CU}^{Z}_{j,\vec{J}}(\theta)+\mathcal{O}(\theta^2)$
          \\[5pt] \hline
              \end{tabular}
    \captionsetup{labelfont=normalfont, textfont=normalfont}
    \caption{\textbf{Compiler rules}, reproduced with permission from \cite{crane2024}. Row 1) $\hat{\Theta}$ is any operator that commutes with $\hat{a}_j$ and $\hat{Z}_j$. Row 2) $K=\lceil\log_2(n_{\mathrm{max}}+1)\rceil$, with $n_{\mathrm{max}}$ is the maximum photon-number cutoff for mode $i$. Rows 2 \& 3) Approximate methods. $\mathrm{CU}^Z_{k,\vec{I}}(\theta)=\exp\left(-i\theta\hat X_k \hat O_{\vec{I}}\right)$, where $\hat{O}_{\vec{I}}$ refers to an operator acting on bosonic modes with indices listed in $\vec{I}$. The superscripts $X$ and $Y$ in $\operatorname{CU}$ 
    denote the qubit axis on which the operator $\hat{U}$ is conditioned. This axis can be controlled by conjugating $\operatorname{CU}^Z$ by single qubit rotations. \label{tab_rules}}
\end{table*}

    \subsubsection{Trotterization}

The \textbf{Trotterization stage} employs the well-known Trotter-Suzuki decomposition, allowing us to rewrite a time-evolution operator $e^{-i(\hat{H}_1 + \hat{H}_2 + \ldots + \hat{H}_n)t}$ as a product of exponentials of individual terms: \begin{equation} e^{-i(\hat{H}_1 + \hat{H}_2 + \cdots + \hat{H}_n)t} \approx \left(\prod_{k} e^{-i\hat{H}_k t/n}\right)^{n}. \end{equation}

For large $n$, the approximation becomes increasingly accurate~\cite{PhysRevX.11.011020}. By isolating each $\hat{H}_k$, we ensure that subsequent passes (Factorization and BCH) can operate on a simpler input: exponentials of single terms rather than complicated multi-term exponentials.

    \subsubsection{Factorization}

The factorization stage consists of two sub-stages: \textbf{density factorization} and \textbf{Pauli factorization}, which originate from Ref.~\cite{crane2024} and we define them in the following. These steps rely on exact mathematical equivalences to rewrite operators without additional approximation error and proceeds from ``left to right'' along the normal ordering, i.e. first number operators are extracted, then qubit operators and finally ladder operators.

\textbf{Density factorization} refers to the step in which $\hat n_i = \hat a_i^\dagger \hat a_i$ operators in the exponential are factored out using the compilation strategy introduced in Ref.~\cite{crane2024}: using a qubit rotation conditioned on the bosonic Fock state, the \textbf{SQR gate} defined in ~\ref{tab:gates}, we make an intermediary transfer of the bosonic density information to a qubit, which is then manipulated, and then the information is transferred back. This encoding allows for the factorization of unitaries involving products of bosonic number operators into sequences of native qubit-boson gates, which usually do not contain such products. This method has a logarithmic overhead in the Fock space dimension. 

More specifically, this method factorizes controlled unitaries of the form \(\operatorname{CU}_{\hat{n}_i,j} = \exp(-i \hat{Z}_{\text{anc}} \hat{n}_i \hat{O}_j)\), where \(\hat{O}_j\) is an Hermitian operator acting on another system (i.e. $j$ is not $i$ and not the ancilla), into a sequence of operations using SQR (Selective Qubit Rotation) gates conditioned on the Fock states of mode \(i\) acting on the ancilla qubit defined in Tab. \ref{tab:gates}. In this method, \(K = \lceil \log_2 n_{\text{max}} \rceil\) is the number of bits required to represent the maximum occupation number \(n_{\text{max}}\) and which will also represent the number of times the subsequence needs to be applied if using a single qubit (in high-Q cavity coupled to transmon hardware, often a single mode is coupled to a single qubit). This decomposition effectively breaks down the controlled unitary into a series of simpler gates.

\textbf{Pauli factorization} In this stage, we systematically isolate qubit operators from bosonic ladder operators by using the second line in Tab.~\ref{tab_rules}. Depending on the operator $\hat{O}_j$ being fully composed of Pauli strings or containing bosonic operators, the decomposition will conjugate a factored term with either a qubit controlled bosonic rotation gate or a CNOT gate resembling the transfer of information onto one of the quantum objects.

    \subsubsection{Baker-Campbell-Hausdorff (BCH)}

Once normal ordering, Trotterization, and factorization have been applied, non-native operators may still remain. 
To break products of raising and lowering operators down, we use the trick introduced in Ref.~\cite{kang2025} and introduce an ancillary qubit and use the BCH formula to approximate $e^{-2i\hat{Z}_k\hat{O}_{\vec{I}}\hat{O}_{\vec{J}}\theta^2}$ as is done in Tab.~\ref{tab_rules}. This trick can be repeatedly applied until $\hat O$ is only a single native gate as discussed in from Tab. \ref{tab:gates}.

\subsection{Generate}
\subsubsection{ISA gate matching}

The native gates into which the results are compiled into belong to the overcomplete ISA provided in Tab. \ref{tab:gates}. The output is a string of numbers and letters. Each line corresponds to a separate qubit-boson gate. The numbers provide the parameters of the gate and the letters provide the operators, all of which when in an exponential give rise to one of the gates defined in the overcomplete ISA in Tab.~\ref{tab:gates}. The compiler's results for the qubit-boson ISA compared to a qubit-only ISA for four paradigmatic Hamiltonians have been plotted in Fig.~\ref{fig:hubbardholstein} and will be discussed in the next section.

A very nice feature of our compiler lies in the fact that it can compile each Hamiltonian Trotter step into qubit-boson gates regardless of the bosonic Hilbert space truncation because it is symbolic and not numeric.

A significant feature of our abstraction is the ability for users to define gate and unitary primitives, offering flexibility in constructing and evolving quantum logic circuits. The integration with Python allows these unitaries to be described as functions, leveraging Python's expressive syntax and functional programming capabilities.

Our system introduces a Circuit object that enables the appending of unitaries into a directed acyclic graph (DAG) structure which introduces a new backend to openfermion, allowing Hamiltonian compilation with OpenFermion as a frontend. This structure effectively represents the desired evolution of the quantum system, allowing for clear visualization and manipulation of the quantum circuit's flow. 

\subsubsection{Optional decomposition into qubit-only ISAs}

If desired, one can take the bosonic unitaries and brute-force numerically decompose them, treating them as arbitrary unitaries. This can be done in QISKit or Circ and is not discussed in this work.

\section{Benchmarking}

\begin{figure}
\centering
\includegraphics[width=0.85\linewidth]{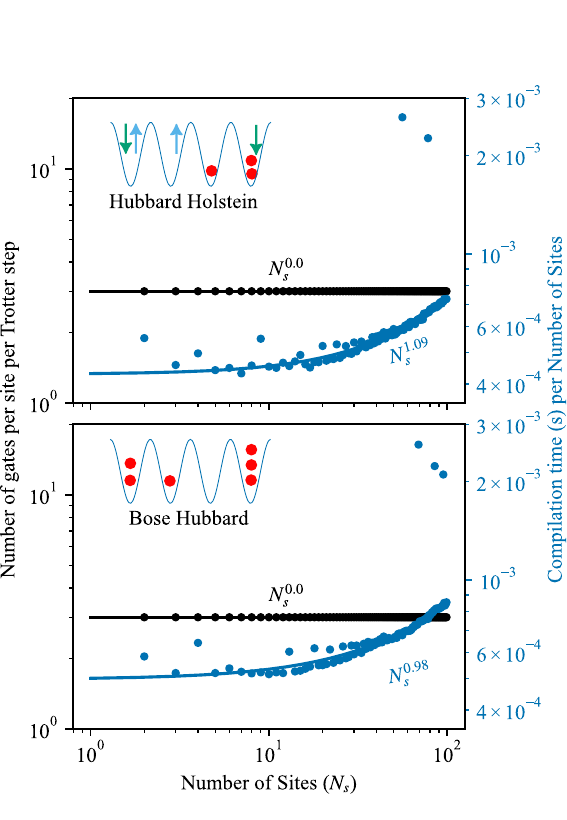}

\caption{\textbf{Compilation time and number of gates as a function of the system size $N_\mathrm{s}$.}  We consider two different benchmark Hamiltonians. Fits to functional form $A N_\mathrm{s}^B + C$ with $A, B$ and $C$ constants are shown with the solid blue (compile time) and black (gate count) along with their asymptotic scaling.
}
\label{fig:hubbardholstein}
\end{figure}

We benchmark the performance of our approach by applying it to the Trotterized time evolution of three of the most prominent quantum systems: the Bose-Hubbard and Hubbard-Holstein Models. 

The Bose-Hubbard model is a toy model for representing interactions of bosonic matter. The simplest form of the Hamiltonian is 
\begin{align}
    \hat{H}_{B.H.} = &\sum_{i, j} t_{i,j}(\hat{b}_{i}^{\dagger}\hat{b}_{j} + \hat{b}_j^{\dagger}\hat{b}_i) + \frac{U}{2}\sum_{i}^{N_\mathrm{s}} \hat{n}_i(\hat{n}_i - 1)\notag\\
    &-\mu \sum_{i}^{N_\mathrm{s}} \hat{n}_{i},
\end{align}
where $\hat{n}_i=\hat{b}_i^{\dagger}\hat{b}_i$ is the number operator, $\hat{b}_i^{\dagger}$ is the bosonic creation operator, $t_{i,j}$ governs the strength of hopping between sites $i$ and $j$, $U$ is the onsite potential, and $\mu$ is a chemical potential term governing the total number of particles in the system.

The Hubbard-Holstein model describes interactions between electrons and vibrational models, known as phonons and is written in Eq.~\ref{eq_HH}.

The Bose-Hubbard \cite{Sawaya:2024qwe,Llodra:2024qhi,Sawaya:2023nmv,Shtanko:2021tsp,Yalouz:2021oyo} and Hubbard-Holstein \cite{Tong:2021rfv,2022PhRvB.106o5158A,Backes:2023lrb,Kumar:2023ubf} have all been used for benchmarking in recent years.
We consider Hamiltonians with both nearest-neighbor hopping on a 1D lattice.

\subsection{Evaluation Metrics}
We use the gate counts and compilation times in seconds for a single Trotter step as our evaluation metrics. For gate count, we plot the scaling of the number of bosonic and and coupled bosonic-qubit gates as a function of the system size for 1D nearest-neighbour hopping.

\subsection{Implementation Details} 
The compiler was implemented entirely in Python (version 3.8) and decompositions were compared with QuTiP to check their correctness. We used n AMD EPYC 92-core CPU and an NVIDIA A100 GPU, with calculations implemented in PyTorch.

\subsection{Data Analysis}

\begin{table}[!ht]
\centering
\caption{Asymptotic scaling of compile time and gate cost for the two target models, where $N_\mathrm{s}$ is the number of sites.}
\label{tab:asymptotics}
\def\arraystretch{1.5}
\begin{tabular}{cccc}
\hline\hline Model & Compile time & Gates\\\hline
Hubbard-Holstein & $\mathcal{O}(N^{2.09}_s)$ & $\mathcal{O}(N_\mathrm{s})$\\
Bose-Hubbard & $\mathcal{O}(N^{1.98}_s)$ & $\mathcal{O}(N_\mathrm{s})$\\

\hline\hline
\end{tabular}
\end{table}

We compiled instances of these models for system sizes ranging from 2 to 100 sites. We found the compilation time and the number of gates per Trotter step to asymptotically fit an ansatz
\begin{align}
    \label{eq:ansatz}
    t = A N_\mathrm{s}^B + C,
\end{align}
where \(A\), \(B\) and \(C\) are fitting parameters and \(N_\mathrm{s}\) is the number of sites in the target model. We show the scaling of the number of gates in Fig.~\ref{fig:hubbardholstein}, extracted from the exponent $B$ obtained from the fit. 

In qubit approaches using the Fock-binary encoding, the number of gates per Trotter step scales quadratically with the number of sites if the boson number cutoff per site is scaled proportional to the system size~\cite{crane2024} (which is what is needed to not introduce additional error for fixed particle density). This is even though there are only a linear number of bosonic terms. By contrast, what is significant in this qubit-boson approach, exemplified by the results of our compiler, is that the number of gates per site is independent on the number of sites.

The total compile time in Table~\ref{tab:asymptotics} is given by the compilation time per number of sites plotted in Fig.~\ref{fig:hubbardholstein}, multiplied by the total number of sites $N_\mathrm{s}$. We find this scales polynomially with the number of sites, with asymptotic behaviors of \(\mathcal{O}(N_\mathrm{s}^{2.09})\) for the Hubbard-Holstein model, and \(\mathcal{O}(N_\mathrm{s}^{1.98})\) for the Bose-Hubbard model.

Furthermore, the scaling behavior—both in terms of gate counts and compile time—demonstrates that the automatic decomposition process introduces no significant quantum gate overhead.

We emphasize that our automatic qubit-boson compilation is manifestly independent of the boson cutoff due to it working on the symbolic level of second-quantized operators. Note that for models including density-density interactions, e.g. a Hamiltonian term $\sum_{i,j} U_{i,j} \hat n_i \hat n_j$, the compilation into the gate set used in this work will yield a logarithmic dependence of the number of gates on the cutoff due to the density factorization stage. This is still exponentially better than the quadratic cutoff dependence found for qubit compilation methods (see e.g. Ref.~\cite{crane2024} and section VIII in Ref.~\cite{liu2024hybrid}).

\section{Conclusion}
In this work, we have introduced an automated quantum compiler for simulating fermion-boson systems with qubit-boson quantum computing hardware. Our compiler accepts arbitrary combinations of fermion-boson-qubit Hamiltonians. Most importantly, for most interactions, it yields results independent of the boson cutoff, therefore manifestly preserving the advantages of qubit-boson hardware.

Our compiler's architecture is structured as a pipeline of mathematical equivalences, enabling efficient symbolic processing of Hamiltonians without resorting to explicitly constructing the Hamiltonian matrix in a basis. The key components of our method include the Trotterization algorithm, exact factorization stages for both Pauli and number operators, and the Baker-Campbell-Hausdorff expansion. These elements work in tandem to decompose complex quantum operations into implementable gate sequences within our universal ISA.

We demonstrated the practicality and efficiency of our compiler through benchmarks on prominent quantum systems, including the Bose-Hubbard Model, and the Hubbard-Holstein Model. The results showed that our compiler scales polynomially with system size in both gate count and compilation time, with no significant overhead introduced by the higher level of abstraction. This confirms the viability of our approach for large-scale quantum simulations and algorithms.

By providing an intuitive interface that leverages second quantization and Pauli operators within a familiar Python environment, we enable users to express complex quantum operations without requiring deep expertise in quantum hardware or theoretical decompositions. 

Future work will focus on further optimizing the compiler's performance and extending its capabilities to accommodate a broader range of quantum systems and hardware architectures.  In particular, in some platforms such as neutral atoms, density-density interactions can be implemented natively, which would remove the logarithmic cutoff dependence of the density factorization stage. 

\section*{Contributions}
ED built compiler framework and performed the majority of simulation runs for data aggregation. 
EG provided insight into models and performance analysis of output of compiler.
SS helped direct research and supervise ED. 
GL supervised ED through course of this work. 
EC developed and designed the project idea, and supervised the project.
All authors contributed to revision of the manuscript.

\section*{Acknowledgments}

EM and EC were supported by the FY24 C2QA Postdoc Seed Funding Award from the Co-design Center for Quantum Advantage. GL and ED were supported in part by the U.S. Department of Energy, Office of Science, Office of Advanced Scientific Computing Research through the Accelerated Research in Quantum Computing Program MACH-Q project, NSF CAREER Award No. CCF-2338773 and ExpandQISE Award No. OSI-2427020. GL is also supported by the Intel Rising Star Award. SS and AL were supported by the U.S. Department of Energy, Office of Science, National Quantum Information Science Research Centers, Co-design Center for Quantum Advantage (C2QA) under contract number DESC0012704, (Basic Energy Sciences). EG was supported by the NASA Academic Mission Services, Contract No. NNA16BD14C and the Intelligent Systems Research and Development-3 (ISRDS-3) Contract 80ARC020D0010 under Co-design Center for Quantum Advantage (C2QA) under Contract No. DE-SC0012704. AS was supported by the U.S. Department of Energy, Office of Science, National Quantum Information Science Research Centers,
Quantum Systems Accelerator (QSA).

\begin{IEEEbiographynophoto}{}
    The authors declare no competing interests?
\end{IEEEbiographynophoto}

\bibliographystyle{IEEEtran}
\bibliography{refs}

\providecommand{\noopsort}[1]{}\providecommand{\singleletter}[1]{#1}%
\begin{thebibliography}{10}
\providecommand{\url}[1]{#1}
\csname url@samestyle\endcsname
\providecommand{\newblock}{\relax}
\providecommand{\bibinfo}[2]{#2}
\providecommand{\BIBentrySTDinterwordspacing}{\spaceskip=0pt\relax}
\providecommand{\BIBentryALTinterwordstretchfactor}{4}
\providecommand{\BIBentryALTinterwordspacing}{\spaceskip=\fontdimen2\font plus
\BIBentryALTinterwordstretchfactor\fontdimen3\font minus \fontdimen4\font\relax}
\providecommand{\BIBforeignlanguage}[2]{{%
\expandafter\ifx\csname l@#1\endcsname\relax
\typeout{** WARNING: IEEEtran.bst: No hyphenation pattern has been}%
\typeout{** loaded for the language `#1'. Using the pattern for}%
\typeout{** the default language instead.}%
\else
\language=\csname l@#1\endcsname
\fi
#2}}
\providecommand{\BIBdecl}{\relax}
\BIBdecl

\bibitem{Cao_2019}
\BIBentryALTinterwordspacing
Y.~Cao, J.~Romero, J.~P. Olson, M.~Degroote, P.~D. Johnson, M.~Kieferová, I.~D. Kivlichan, T.~Menke, B.~Peropadre, N.~P.~D. Sawaya, S.~Sim, L.~Veis, and A.~Aspuru-Guzik, ``Quantum chemistry in the age of quantum computing,'' \emph{Chemical Reviews}, vol. 119, no.~19, pp. 10\,856--10\,915, 2019, pMID: 31469277. [Online]. Available: \url{https://doi.org/10.1021/acs.chemrev.8b00803}
\BIBentrySTDinterwordspacing

\bibitem{TILLY20221}
\BIBentryALTinterwordspacing
J.~Tilly, H.~Chen, S.~Cao, D.~Picozzi, K.~Setia, Y.~Li, E.~Grant, L.~Wossnig, I.~Rungger, G.~H. Booth, and J.~Tennyson, ``The variational quantum eigensolver: A review of methods and best practices,'' \emph{Physics Reports}, vol. 986, pp. 1--128, 2022, the Variational Quantum Eigensolver: a review of methods and best practices. [Online]. Available: \url{https://www.sciencedirect.com/science/article/pii/S0370157322003118}
\BIBentrySTDinterwordspacing

\bibitem{Sawaya:2023nmv}
\BIBentryALTinterwordspacing
N.~P. Sawaya, D.~Marti-Dafcik, Y.~Ho, D.~P. Tabor, D.~E.~B. Neira, A.~B. Magann, S.~Premaratne, P.~Dubey, A.~Matsuura, N.~Bishop, W.~A.~d. Jong, S.~Benjamin, O.~Parekh, N.~Tubman, K.~Klymko, and D.~Camps, ``Ham{L}ib: {A} library of {H}amiltonians for benchmarking quantum algorithms and hardware,'' \emph{{Quantum}}, vol.~8, p. 1559, Dec. 2024. [Online]. Available: \url{https://doi.org/10.22331/q-2024-12-11-1559}
\BIBentrySTDinterwordspacing

\bibitem{Bauer:2022hpo}
\BIBentryALTinterwordspacing
C.~W. Bauer \emph{et~al.}, ``{Quantum Simulation for High-Energy Physics},'' \emph{PRX Quantum}, vol.~4, no.~2, p. 027001, 2023. [Online]. Available: \url{https://doi.org/10.1103/PRXQuantum.4.027001}
\BIBentrySTDinterwordspacing

\bibitem{Alam:2022crs}
\BIBentryALTinterwordspacing
M.~S. Alam \emph{et~al.}, ``{Quantum computing hardware for HEP algorithms and sensing},'' in \emph{{Snowmass 2021}}, 4 2022. [Online]. Available: \url{https://lss.fnal.gov/archive/2022/pub/fermilab-pub-22-260-sqms.pdf}
\BIBentrySTDinterwordspacing

\bibitem{blekos2024review}
\BIBentryALTinterwordspacing
K.~Blekos, D.~Brand, A.~Ceschini, C.-H. Chou, R.-H. Li, K.~Pandya, and A.~Summer, ``A review on quantum approximate optimization algorithm and its variants,'' \emph{Physics Reports}, vol. 1068, pp. 1--66, 2024. [Online]. Available: \url{https://doi.org/10.1016/j.physrep.2024.03.002}
\BIBentrySTDinterwordspacing

\bibitem{crane2024}
\BIBentryALTinterwordspacing
E.~Crane, K.~C. Smith, T.~Tomesh, A.~Eickbusch, J.~M. Martyn, S.~Kühn, L.~Funcke, M.~A. DeMarco, I.~L. Chuang, N.~Wiebe, A.~Schuckert, and S.~M. Girvin, ``Hybrid oscillator-qubit quantum processors: Simulating fermions, bosons, and gauge fields,'' 2024. [Online]. Available: \url{https://arxiv.org/abs/2409.03747}
\BIBentrySTDinterwordspacing

\bibitem{liu2024hybrid}
\BIBentryALTinterwordspacing
Y.~Liu, S.~Singh, K.~C. Smith, E.~Crane, J.~M. Martyn, A.~Eickbusch, A.~Schuckert, R.~D. Li, J.~Sinanan-Singh, M.~B. Soley \emph{et~al.}, ``Hybrid oscillator-qubit quantum processors: Instruction set architectures, abstract machine models, and applications,'' 2024. [Online]. Available: \url{https://arxiv.org/abs/2407.10381}
\BIBentrySTDinterwordspacing

\bibitem{gonzalez-cuadra_fermionic_2023}
\BIBentryALTinterwordspacing
D.~González-Cuadra, D.~Bluvstein, M.~Kalinowski, R.~Kaubruegger, N.~Maskara, P.~Naldesi, T.~V. Zache, A.~M. Kaufman, M.~D. Lukin, H.~Pichler, B.~Vermersch, J.~Ye, and P.~Zoller, ``Fermionic quantum processing with programmable neutral atom arrays,'' \emph{Proceedings of the National Academy of Sciences}, vol. 120, no.~35, p. e2304294120, Aug. 2023, publisher: Proceedings of the National Academy of Sciences. [Online]. Available: \url{https://www.pnas.org/doi/full/10.1073/pnas.2304294120}
\BIBentrySTDinterwordspacing

\bibitem{Schuckert_2024}
\BIBentryALTinterwordspacing
A.~Schuckert, E.~Crane, A.~V. Gorshkov, M.~Hafezi, and M.~J. Gullans, ``{Fermion-qubit fault-tolerant quantum computing},'' 11 2024. [Online]. Available: \url{https://arxiv.org/abs/2411.08955}
\BIBentrySTDinterwordspacing

\bibitem{dawson2005solovaykitaevalgorithm}
\BIBentryALTinterwordspacing
C.~M. Dawson and M.~A. Nielsen, ``The solovay-kitaev algorithm,'' 2005. [Online]. Available: \url{https://arxiv.org/abs/quant-ph/0505030}
\BIBentrySTDinterwordspacing

\bibitem{childs_theory_2021}
\BIBentryALTinterwordspacing
A.~M. Childs, Y.~Su, M.~C. Tran, N.~Wiebe, and S.~Zhu, ``Theory of {Trotter} {Error} with {Commutator} {Scaling},'' \emph{Physical Review X}, vol.~11, no.~1, p. 011020, Feb. 2021, publisher: American Physical Society. [Online]. Available: \url{https://link.aps.org/doi/10.1103/PhysRevX.11.011020}
\BIBentrySTDinterwordspacing

\bibitem{mcclean2019openfermionelectronicstructurepackage}
\BIBentryALTinterwordspacing
J.~R. McClean, K.~J. Sung, I.~D. Kivlichan, Y.~Cao, C.~Dai, E.~S. Fried, C.~Gidney, B.~Gimby, P.~Gokhale, T.~Häner, T.~Hardikar, V.~Havlíček, O.~Higgott, C.~Huang, J.~Izaac, Z.~Jiang, X.~Liu, S.~McArdle, M.~Neeley, T.~O'Brien, B.~O'Gorman, I.~Ozfidan, M.~D. Radin, J.~Romero, N.~Rubin, N.~P.~D. Sawaya, K.~Setia, S.~Sim, D.~S. Steiger, M.~Steudtner, Q.~Sun, W.~Sun, D.~Wang, F.~Zhang, and R.~Babbush, ``Openfermion: The electronic structure package for quantum computers,'' 2019. [Online]. Available: \url{https://arxiv.org/abs/1710.07629}
\BIBentrySTDinterwordspacing

\bibitem{qiskit2024}
A.~Javadi-Abhari, M.~Treinish, K.~Krsulich, C.~J. Wood, J.~Lishman, J.~Gacon, S.~Martiel, P.~D. Nation, L.~S. Bishop, A.~W. Cross, B.~R. Johnson, and J.~M. Gambetta, ``Quantum computing with {Q}iskit,'' 2024.

\bibitem{Killoran_2019}
\BIBentryALTinterwordspacing
N.~Killoran, J.~Izaac, N.~Quesada, V.~Bergholm, M.~Amy, and C.~Weedbrook, ``Strawberry fields: A software platform for photonic quantum computing,'' \emph{Quantum}, vol.~3, p. 129, Mar. 2019. [Online]. Available: \url{http://dx.doi.org/10.22331/q-2019-03-11-129}
\BIBentrySTDinterwordspacing

\bibitem{bernardini2023quantumcomputingtrappedions}
\BIBentryALTinterwordspacing
F.~Bernardini, A.~Chakraborty, and C.~Ordóñez, ``Quantum computing with trapped ions: a beginner's guide,'' 2023. [Online]. Available: \url{https://arxiv.org/abs/2303.16358}
\BIBentrySTDinterwordspacing

\bibitem{google2023suppressing}
\BIBentryALTinterwordspacing
G.~Q. AI, ``Suppressing quantum errors by scaling a surface code logical qubit,'' \emph{Nature}, vol. 614, pp. 676--681, 2023. [Online]. Available: \url{https://doi.org/10.1038/s41586-022-05434-1}
\BIBentrySTDinterwordspacing

\bibitem{bluvstein2024logical}
\BIBentryALTinterwordspacing
D.~Bluvstein, S.~J. Evered, A.~A. Geim \emph{et~al.}, ``Logical quantum processor based on reconfigurable atom arrays,'' \emph{Nature}, vol. 626, pp. 58--65, 2024. [Online]. Available: \url{https://doi.org/10.1038/s41586-023-06927-3}
\BIBentrySTDinterwordspacing

\bibitem{cirq_developers_2024_11398048}
\BIBentryALTinterwordspacing
C.~Developers, ``Cirq,'' May 2024. [Online]. Available: \url{https://doi.org/10.5281/zenodo.11398048}
\BIBentrySTDinterwordspacing

\bibitem{PhysRevA.97.062311}
\BIBentryALTinterwordspacing
T.~Kalajdzievski, C.~Weedbrook, and P.~Rebentrost, ``Continuous-variable gate decomposition for the bose-hubbard model,'' \emph{Phys. Rev. A}, vol.~97, p. 062311, Jun 2018. [Online]. Available: \url{https://link.aps.org/doi/10.1103/PhysRevA.97.062311}
\BIBentrySTDinterwordspacing

\bibitem{javadiabhari2024quantumcomputingqiskit}
\BIBentryALTinterwordspacing
A.~Javadi-Abhari, M.~Treinish, K.~Krsulich, C.~J. Wood, J.~Lishman, J.~Gacon, S.~Martiel, P.~D. Nation, L.~S. Bishop, A.~W. Cross, B.~R. Johnson, and J.~M. Gambetta, ``Quantum computing with qiskit,'' 2024. [Online]. Available: \url{https://arxiv.org/abs/2405.08810}
\BIBentrySTDinterwordspacing

\bibitem{decker2025kernpilercompileroptimizationquantum}
\BIBentryALTinterwordspacing
E.~Decker, L.~Goetz, E.~McKinney, E.~Gustafson, J.~Zhou, Y.~Liu, A.~K. Jones, A.~Li, A.~Schuckert, S.~Stein, E.~Crane, and G.~Li, ``Kernpiler: Compiler optimization for quantum hamiltonian simulation with partial trotterization,'' 2025. [Online]. Available: \url{https://arxiv.org/abs/2504.07214}
\BIBentrySTDinterwordspacing

\bibitem{Stavenger:2022wzz}
\BIBentryALTinterwordspacing
T.~J. Stavenger, E.~Crane, K.~C. Smith, C.~T. Kang, S.~M. Girvin, and N.~Wiebe, ``C2qa - bosonic qiskit,'' in \emph{2022 IEEE High Performance Extreme Computing Conference (HPEC)}, 2022, pp. 1--8. [Online]. Available: \url{https://ieeexplore.ieee.org/abstract/document/9926318}
\BIBentrySTDinterwordspacing

\bibitem{Mato:2024hnd}
\BIBentryALTinterwordspacing
K.~Mato, M.~Ringbauer, L.~Burgholzer, and R.~Wille, ``Mqt qudits: A software framework for mixed-dimensional quantum computing,'' \emph{arXiv preprint arXiv:2410.02854}, 2024. [Online]. Available: \url{https://arxiv.org/abs/2410.02854}
\BIBentrySTDinterwordspacing

\bibitem{Peng:2023fmd}
\BIBentryALTinterwordspacing
Y.~Peng, J.~Young, P.~Liu, and X.~Wu, ``Simuq: A framework for programming quantum hamiltonian simulation with analog compilation,'' \emph{Proc. ACM Program. Lang.}, vol.~8, no. POPL, Jan. 2024. [Online]. Available: \url{https://doi.org/10.1145/3632923}
\BIBentrySTDinterwordspacing

\bibitem{zhou:2024bosehedral}
\BIBentryALTinterwordspacing
J.~Zhou, Y.~Liu, Y.~Shi, A.~Javadi-Abhari, and G.~Li, ``Bosehedral: Compiler optimization for bosonic quantum computing,'' in \emph{2024 ACM/IEEE 51st Annual International Symposium on Computer Architecture (ISCA)}, 2024, pp. 261--276. [Online]. Available: \url{https://ieeexplore.ieee.org/abstract/document/10609638}
\BIBentrySTDinterwordspacing

\bibitem{kang2025}
\BIBentryALTinterwordspacing
C.~Kang, M.~Soley, E.~Crane, S.~M. Girvin, and N.~Wiebe, ``Leveraging hamiltonian simulation techniques to compile operations on bosonic devices,'' \emph{Journal of Physics A: Mathematical and Theoretical}, 2025. [Online]. Available: \url{http://iopscience.iop.org/article/10.1088/1751-8121/adb5df}
\BIBentrySTDinterwordspacing

\bibitem{Sefi_2011}
\BIBentryALTinterwordspacing
S.~Sefi and P.~Van~Loock, ``How to decompose arbitrary continuous-variable quantum operations,'' \emph{Physical review letters}, vol. 107, no.~17, p. 170501, 2011. [Online]. Available: \url{http://dx.doi.org/10.1103/PhysRevLett.107.170501}
\BIBentrySTDinterwordspacing

\bibitem{crane2021}
\BIBentryALTinterwordspacing
E.~Crane, ``Quantum computation and simulation in silicon donors: from optically-controlled entangling gates to the hubbard model,'' Ph.D. dissertation, UCL (University College London), 2021. [Online]. Available: \url{https://discovery.ucl.ac.uk/id/eprint/10140860/}
\BIBentrySTDinterwordspacing

\bibitem{fermi-hubbard}
\BIBentryALTinterwordspacing
K.~H\'emery, K.~Ghanem, E.~Crane, S.~L. Campbell, J.~M. Dreiling, C.~Figgatt, C.~Foltz, J.~P. Gaebler, J.~Johansen, M.~Mills, S.~A. Moses, J.~M. Pino, A.~Ransford, M.~Rowe, P.~Siegfried, R.~P. Stutz, H.~Dreyer, A.~Schuckert, and R.~Nigmatullin, ``Measuring the loschmidt amplitude for finite-energy properties of the fermi-hubbard model on an ion-trap quantum computer,'' \emph{PRX Quantum}, vol.~5, p. 030323, Aug 2024. [Online]. Available: \url{https://link.aps.org/doi/10.1103/PRXQuantum.5.030323}
\BIBentrySTDinterwordspacing

\bibitem{Nigmatullin:2024mbq}
\BIBentryALTinterwordspacing
R.~Nigmatullin, K.~Hemery, K.~Ghanem, S.~Moses, D.~Gresh, P.~Siegfried, M.~Mills, T.~Gatterman, N.~Hewitt, E.~Granet, and H.~Dreyer, ``Experimental demonstration of break-even for the compact fermionic encoding,'' 2024. [Online]. Available: \url{https://arxiv.org/abs/2409.06789}
\BIBentrySTDinterwordspacing

\bibitem{kivlichan_quantum_2018}
\BIBentryALTinterwordspacing
I.~D. Kivlichan, J.~McClean, N.~Wiebe, C.~Gidney, A.~Aspuru-Guzik, G.~K.-L. Chan, and R.~Babbush, ``Quantum simulation of electronic structure with linear depth and connectivity,'' \emph{Phys. Rev. Lett.}, vol. 120, p. 110501, Mar 2018. [Online]. Available: \url{https://link.aps.org/doi/10.1103/PhysRevLett.120.110501}
\BIBentrySTDinterwordspacing

\bibitem{bravyi2002fermionic}
\BIBentryALTinterwordspacing
S.~B. Bravyi and A.~Y. Kitaev, ``Fermionic quantum computation,'' \emph{Annals of Physics}, vol. 298, no.~1, pp. 210--226, 2002. [Online]. Available: \url{https://www.sciencedirect.com/science/article/pii/S0003491602962548}
\BIBentrySTDinterwordspacing

\bibitem{seeley2012bravyi}
\BIBentryALTinterwordspacing
J.~T. Seeley, M.~J. Richard, and P.~J. Love, ``The bravyi-kitaev transformation for quantum computation of electronic structure,'' \emph{The Journal of Chemical Physics}, vol. 137, no.~22, p. 224109, 12 2012. [Online]. Available: \url{https://doi.org/10.1063/1.4768229}
\BIBentrySTDinterwordspacing

\bibitem{PhysRevX.11.011020}
\BIBentryALTinterwordspacing
A.~M. Childs, Y.~Su, M.~C. Tran, N.~Wiebe, and S.~Zhu, ``Theory of trotter error with commutator scaling,'' \emph{Phys. Rev. X}, vol.~11, p. 011020, Feb 2021. [Online]. Available: \url{https://link.aps.org/doi/10.1103/PhysRevX.11.011020}
\BIBentrySTDinterwordspacing

\bibitem{Sawaya:2024qwe}
\BIBentryALTinterwordspacing
N.~P. Sawaya, D.~Camps, N.~M. Tubman, G.~M. Rotskoff, and R.~LaRose, ``Non-clifford diagonalization for measurement shot reduction in quantum expectation value estimation,'' 2024. [Online]. Available: \url{https://arxiv.org/abs/2408.11898}
\BIBentrySTDinterwordspacing

\bibitem{Llodra:2024qhi}
\BIBentryALTinterwordspacing
G.~Llodrà, P.~Mujal, R.~Zambrini, and G.~L. Giorgi, ``Quantum reservoir computing in atomic lattices,'' \emph{Chaos, Solitons \& Fractals}, vol. 195, p. 116289, 2025. [Online]. Available: \url{https://www.sciencedirect.com/science/article/pii/S0960077925003029}
\BIBentrySTDinterwordspacing

\bibitem{Shtanko:2021tsp}
\BIBentryALTinterwordspacing
O.~Shtanko and R.~Movassagh, ``Preparing thermal states on noiseless and noisy programmable quantum processors,'' 2023. [Online]. Available: \url{https://arxiv.org/abs/2112.14688}
\BIBentrySTDinterwordspacing

\bibitem{Yalouz:2021oyo}
\BIBentryALTinterwordspacing
S.~Yalouz, B.~Senjean, F.~Miatto, and V.~Dunjko, ``Encoding strongly-correlated many-boson wavefunctions on a photonic quantum computer: application to the attractive {B}ose-{H}ubbard model,'' \emph{{Quantum}}, vol.~5, p. 572, Nov. 2021. [Online]. Available: \url{https://doi.org/10.22331/q-2021-11-08-572}
\BIBentrySTDinterwordspacing

\bibitem{Tong:2021rfv}
\BIBentryALTinterwordspacing
Y.~Tong, V.~V. Albert, J.~R. McClean, J.~Preskill, and Y.~Su, ``Provably accurate simulation of gauge theories and bosonic systems,'' \emph{{Quantum}}, vol.~6, p. 816, Sep. 2022. [Online]. Available: \url{https://doi.org/10.22331/q-2022-09-22-816}
\BIBentrySTDinterwordspacing

\bibitem{2022PhRvB.106o5158A}
\BIBentryALTinterwordspacing
R.~J. Anderson, C.~J.~C. Scott, and G.~H. Booth, ``Full configuration interaction quantum monte carlo for coupled electron-boson systems and infinite spaces,'' \emph{Phys. Rev. B}, vol. 106, p. 155158, Oct 2022. [Online]. Available: \url{https://link.aps.org/doi/10.1103/PhysRevB.106.155158}
\BIBentrySTDinterwordspacing

\bibitem{Backes:2023lrb}
\BIBentryALTinterwordspacing
S.~Backes, Y.~Murakami, S.~Sakai, and R.~Arita, ``Dynamical mean-field theory for the hubbard-holstein model on a quantum device,'' \emph{Phys. Rev. B}, vol. 107, p. 165155, Apr 2023. [Online]. Available: \url{https://link.aps.org/doi/10.1103/PhysRevB.107.165155}
\BIBentrySTDinterwordspacing

\bibitem{Kumar:2023ubf}
\BIBentryALTinterwordspacing
S.~Kumar, N.~N. Hegade, A.-M. Visuri, B.~A. Bhargava, J.~F. Hernandez, E.~Solano, F.~Albarr{\'a}n-Arriagada, and G.~A. Barrios, ``Digital-analog quantum computing of fermion-boson models in superconducting circuits,'' \emph{npj Quantum Information}, vol.~11, no.~1, p.~43, 2025. [Online]. Available: \url{https://www.nature.com/articles/s41534-025-01001-4}
\BIBentrySTDinterwordspacing

\end{thebibliography}

\end{document}